\documentclass[aps,eqsecnum,pra, twocolumn]{revtex4}
\usepackage{graphics,graphicx}
\usepackage{amsmath}
\usepackage{color}
\usepackage{amssymb,latexsym,mathrsfs}
\usepackage{hyperref}
\usepackage{morefloats}
\def\bea{\begin{eqnarray}}
\def\eea{\end{eqnarray}}
\def\ba{\begin{array}}
\def\ea{\end{array}}

\def\ket{\rangle}
\def\bra{\langle}
\def\beq{\begin{equation}}
\def\eeq{\end{equation}}

\begin{document}

\title{Evolution of coherence and non-classicality under global environmental interaction}

\author{Samyadeb Bhattacharya$^{1,2}$ \footnote{sbh.phys@gmail.com}, Subhashish Banerjee$^{3}$ \footnote{subhashish@iitj.ac.in},
 Arun Kumar Pati$^{1}$ \footnote{akpati@hri.res.in}}
\affiliation{$^{1}$Harish-Chandra Research Institute, Chhatnag Road, Jhunsi, Allahabad - 211019, India \\}
\affiliation{$^{2}$ S. N. Bose National Centre for Basic Sciences, Block JD, Sector III, Salt Lake, Kolkata 700 098, India}
\affiliation{$^{3}$Indian Institute of Technology Rajasthan, Jodhpur, India\\}


\begin{abstract}

\noindent A master equation has been constructed for a global system-bath interaction both in the absence as well as presence of non-Markovian noise. For the memoryless case, it has been exactly solved for a paradigmatic class of two qubit states in high and zero temperature thermal environment. For the non-Markovian model, it has been solved for zero temperature bath. The evolution of quantum coherence and entanglement has been observed in presence of the above mentioned interactions. We show that the
global part of the system-bath interaction compensates for the decoherence, resulting in slow down of coherence and entanglement decay. For an appropriately defined limiting case, both coherence and entanglement show freezing behaviour for the high temperature bath. In case of zero temperature bath, the mentioned interaction not only stabilizes the non-classical correlations, but also enhances them for a finite period. For the memory dependent case, we have seen that the global interaction enhances the back-flow of information from environment to the system, as it enhances the regeneration of coherence and entanglement. Also we have studied the generation of Quantum Fisher information by the mentioned process. An intuitive measure of non-classicality based on non-commutativity of quantum states has been considered. Bounds on generated quantum Fisher information has been found in terms of quantumness and coherence. This gives us a novel understanding of Quantum Fisher information as a measure of non-classicality.
\vspace{0.5cm}


\end{abstract}

\maketitle

\section{Introduction}

The main objective of the theory of open quantum systems is to develop a comprehensive description of various kind of interactions
of the system with its ambient environment and their effect on the dynamics of the system of interest \cite{breuer,weiss}. Any realistic quantum system
is bound to get affected by its environment and therefore the dynamical features of open quantum systems are particularly important from
the practical perspective. Due to the advent of quantum technologies such as quantum communication \cite{teleport,comm1,comm2}, quantum cryptography \cite{nelson, crypto}, there has
been an upsurge of interest in the application of several techniques of open quantum systems. Markovian dynamics of open systems have been extensively studied
and implemented successfully in a wide variety of problems \cite{breuer, gp, sgad, qw, wigner, tomo}. \\

\noindent However, due to impressive
developments in the experimentation with quantum systems and their control \cite{control}, it has been realized that open quantum systems do not generally behave
within the domain of Markovian dynamics \cite{sg, qbm}. Particularly for the case of systems constituting more than one interacting sub-systems, where the
interactions between the parties may be comparable to the interaction strength of the coupling to the bath, it is natural for the dynamics to
deviate from Markovian behavior \cite{fleming}. Effects of non-Markovian structured environment on quantum coherence and correlation has been theoretically and experimentally studied \cite{franco1,franco2,franco3,franco4,franco5}.
A number of measures of deviation from Markovianity have been proposed recently, where
``Non-Markovianity" is recognized as the departure from monotonic behavior of certain measures under strictly Markovian dynamics
\cite{wolf,breuer2,laine,rivas,raja,breuer3,s1,s2,s3}. Generally, a quantitative measure, which shows monotonic behavior under Markovian dynamics
is addressed and departure from that monotonicity is taken to be a signature of non-Markovianity. For example, we can consider the dynamics of
entanglement in an open system scenario. Since local trace preserving completely positive maps do not increase the amount of entanglement, evidently
it can be surmised that the entanglement with an ancillary system decays monotonically within Markovian regime. Thus, for example, in Ref. \cite{rivas}
the dynamics of entanglement was used as a signature to identify the non-Markovian characteristics of a system-environment interaction. \\

\noindent In recent times, the theory of quantum coherence as a resource has attracted much attention \cite{plenio,winter1,uttam1,uttam2,uttam3,manab1,rana1,strelsov}. 
Coherence plays a central
role in quantum mechanics enabling operations or tasks which are impossible within the regime of classical mechanics. Coherence underlies the
non-classical behavior of a quantum system, like entanglement. In Ref. \cite{plenio} several measures, like the $l_1$ norm and the relative
entropy were introduced to characterize the coherence of a quantum system. The valid measures of coherence should not increase under allowed incoherent operations.
Based on this, it can be inferred that the measures of quantum coherence are monotone under Markovian dynamics. Hence any deviation
from the montonicity of coherence measure can also be taken as a signature of deviation from Markovian dynamics \cite{titas,ning}. The central theme of our work revolves
around this specific issue of the dynamical behavior of coherence monotones under a specific system-environment interaction which is not strictly
Markovian. Usually for the case of Markovian dynamics of a composite system, the environment acts locally on each of the parties. For the bipartite case, considered here,
the environment is globally interacting with the system. This enables us to consider two parts of the non-unitary evolution. One is of
course the local Lindbladian and the other is an interactive part which is essentially causing the deviation from Markovian evolution. Here we have taken
a two qubit atomic system interacting with a Harmonic oscillator bath. A similar model with a squeezed thermal bath has been studied earlier for estimating
the entanglement dynamics of the atomic system \cite{banerjee,tanas}. Based on the above mentioned model, the first
part of our work is to present an analytical estimation of the dynamics of coherence, for the purpose of characterization of the deviation from
monotonicity and to find the conditions under which such deviation occurs. After that we will extend our study to calculate the exact expression of
concurrence for a special class of two qubit X states in order to observe the dynamics of entanglement in our proposed system-bath interaction. \\

\noindent Next, we consider the generation of quantum Fisher information by the global environmental interaction. The Fisher information is a
measure of intrinsic accuracy in statistical ensemble theory \cite{fisher,crammer-rao}. Quantum generalization of the Fisher information
has also been introduced \cite{helstrom,holevo}, which can be considered as a measure of non-classicality for quantum system \cite{hall, sbomkar}. A measure of quantumness has been proposed in a recent work \cite{vedral}, based on the non-commutativity of quantum states. We have shown that the Quantum Fisher information is lower bounded by the above mentioned quantumness with a factor of 1/4 and upper bounded by the $l_1$ norm of coherence with a factor of 1/2. This gives us an intuitive understanding of Quantum Fisher information as a measure of non-classicality. We have estimated the amount of quantumness and Fisher information that can be generated by the global environmental interaction.
Then, we extend our study by considering the memory effect of environmental interaction. We have used a non-Markovian noise model to generalize the global master equation for memory dependent system-bath interactions. We also solve the master equation for two
qubit X states to obtain analytical expressions for the $l_1$ norm of coherence and concurrence and study their dynamics. Further we also investigate the generation of quantumness in the memory dependent environmental interaction.\\

\noindent The organisation of the paper is as follows. In Section II, we will construct the master equation of the concerned system and obtain analytical solution for a particular class of density
matrices. In Sections III and  IV,  we  discuss the dynamics of quantum coherence, concurrence  quantum Fisher information and quantumness, respectively. In Section V we generalize
the master equation to incorporate memory dependent environmental interactions. We then present our conclusions in Section VI.

\section{Dynamics of composite quantum system in global environmental interaction}
The dynamics of the reduced density matrix of the system of interest, undergoing a completely positive (CP) evolution, can be expressed in terms of
the Kraus operator sum representation (OSR) as

\beq\label{secIa}
\rho(t) = \sum_{i} M_i(t)\rho(0)M_i^{\dagger}(t).
\eeq

\noindent The Kraus operators $M_i$ must satisfy the trace preserving condition, i.e.,

\beq\label{secIb}
\sum_i M_i^{\dagger}(t)M_i(t) = \mathbb{I}.
\eeq
The Kraus operator sum representation governs a CP evolution.
The Markovian master equation can be constructed from the Kraus representation \cite{preskill}. For a small time interval $\delta t$, if we consider
\beq\label{secIc}
\begin{array}{ll}
M_i(t) \approx \sqrt{\delta t}L_i, ~~~~~ \forall~ i \neq 0\\
M_0(t) \approx \mathbb{I} - \frac{\delta t}{2}\sum_{i\neq 0} L_{i}^{\dagger}L_{i},
\end{array}
\eeq

then it can be seen from equation (\ref{secIa}), that

\beq\label{secId}
\frac{\rho(t)-\rho(0)}{\delta t} = \sum_i \left[ L_i\rho L_i^{\dagger}-\frac{1}{2}\left( L_i^{\dagger}L_i \rho +
\rho L_i^{\dagger}L_i \right) \right].
\eeq

Now if we consider $L_i=\sqrt{\gamma_i}A_i$ , then for $\delta t\rightarrow 0$, we get the differential equation

\beq\label{secI1}
\frac{d\rho}{dt} = \sum_i \gamma_i \left[ A_i\rho A_i^{\dagger}-\frac{1}{2}\left( A_i^{\dagger}A_i \rho + \rho A_i^{\dagger}A_i \right) \right],
\eeq

\noindent where $\gamma_i$s are the decay parameters. This is basically the usual Markovian master equation. Following Ref. \cite{lider}, for a fixed operator
basis $\{ A_{\alpha} \}_{\alpha=0}^n$, with $A_0=\mathbb{I}$, we can define $L_i=\sum_{\alpha=0}^{n} b_{i\alpha} A_{\alpha}$. Using equation
(\ref{secIc}), it can be seen that the master equation becomes
\beq\label{secI1b}
\frac{d\rho}{dt} = \sum_{\alpha,\beta} \gamma_{\alpha\beta}
\left[ A_{\alpha}\rho A_{\beta}^{\dagger}-\frac{1}{2}\left( A_{\alpha}^{\dagger}A_{\beta} \rho + \rho A_{\alpha}^{\dagger}A_{\beta} \right) \right],
\eeq

\noindent where $\{ \gamma_{\alpha\beta}=\sum_i b_{i\alpha}b_{i\beta}^{*} \}$ represents a coefficient matrix, called as the damping basis. The positivity of this kind of dynamical map depends
on the positivity of the $\{ \gamma_{\alpha\beta} \}$ matrix \cite{lider}. If all the eigenvalues of the damping basis matrix are positive, then
the map is completely positive. \\

\noindent Here we are dealing with a two qubit atomic system. Let us first begin by writing down the master equation for such systems, which is of
the form (\ref{secI1})

\begin{widetext}
\beq\label{secI2}
\frac{d\rho}{dt} = \sum_{i=1,2}\gamma_i (N+1) \left( \sigma_i^{-}\rho\sigma_i^{+} -\frac{1}{2}\left( \sigma_i^{+}\sigma_i^{-}\rho + \rho\sigma_i^{+}\sigma_i^{-} \right) \right)
+ \sum_{i=1,2}\gamma_i N \left( \sigma_i^{+}\rho\sigma_i^{-} -\frac{1}{2}\left( \sigma_i^{-}\sigma_i^{+}\rho + \rho\sigma_i^{-}\sigma_i^{+} \right) \right),
\eeq
\end{widetext}

where

\beq\label{secI2a}
N = \frac{1}{\exp (\frac{\hbar\omega}{k_B T})-1},
\eeq

is the Planck distribution function and

\beq\label{secI6}
\begin{array}{ll}
\sigma_1^{-} = \sigma^{-}\otimes \mathbb{I}~~;~~\sigma_1^{+} = \sigma^{+}\otimes \mathbb{I},\\
\sigma_2^{-} = \mathbb{I} \otimes \sigma^{-}~~;~~\sigma_2^{+} = \mathbb{I} \otimes \sigma^{+},
\end{array}
\eeq

\noindent are the dipole lowering and raising operators acting locally on the two parties. \\
Let us now generalize this master equation for the two qubit system following \cite{banerjee}, where the environment is modelled as a thermal radiation field, which is interacting with the system
in a global way. This is achieved by a coupling dependent upon the qubit position $\bf{r}_n$.
The effective Hamiltonian of the two qubit system can be expressed as 

\beq\label{secI7}
H_S = \sum_{n=1,2} \hbar\omega_n \sigma^{z}_n + H_{int},
\eeq

with 

\beq\label{secI7a}
H_{int}=\hbar\left( \Omega_{21} \sigma^{-}\otimes\sigma^{+}+ \Omega_{12}\sigma^{+}\otimes\sigma^{-}\right)
\eeq

where
\begin{widetext}
\beq\label{secI8}
\Omega_{ij} = \frac{3}{4} \sqrt{\gamma_i\gamma_j} \left[ -(1-(\hat{\mu}.\hat{r}_{ij})^2)\frac{\cos k_0 r_{ij}}{k_0 r_{ij}}+
(1-3(\hat{\mu}.\hat{r}_{ij})^2) \left[ \frac{\sin k_0 r_{ij}}{(k_0 r_{ij})^2}+\frac{\cos k_0 r_{ij}}{(k_0 r_{ij})^3} \right] \right].
\eeq
\end{widetext}

\noindent Here $\hat{\mu}=\hat{\mu}_1=\hat{\mu}_2$ are the dipole moment operators and $\hat{r_{ij}}$ are the unit vectors along the atomic transition dipole moments with $\bf{r}_{ij}=\bf{r}_i-\bf{r}_j$.
Also $k_0=\omega_0/c$, where $\omega_0=(\omega_1+\omega_2)/2$ and $\gamma_i=\omega_i^3\mu_i^2/3\pi\varepsilon\hbar c^3$ is the spontaneous emission rate.
Following Ref. \cite{banerjee}, the non-unitary part of the global master equation can be written as
\begin{widetext}
\beq\label{secI9}
\frac{d\rho}{dt} = \frac{i}{\hbar}[\rho,H_S]+ \sum_{i,j=1,2}\gamma_{ij} (N+1) \left( \sigma_i^{-}\rho\sigma_j^{+} -\frac{1}{2}\left( \sigma_i^{+}\sigma_j^{-}\rho + \rho\sigma_i^{+}\sigma_j^{-} \right) \right)
+ \sum_{i,j=1,2}\gamma_{ij} N \left( \sigma_i^{+}\rho\sigma_j^{-} -\frac{1}{2}\left( \sigma_i^{-}\sigma_j^{+}\rho + \rho\sigma_i^{-}\sigma_j^{+} \right) \right),
\eeq
\end{widetext}

where

\beq\label{secI10}
\begin{array}{ll}
\gamma_{ij} = \sqrt{\gamma_i\gamma_j}  \large{a} (k_0r_{ij})~ ; ~~~ \forall i \neq j\\
\gamma_i    = \frac{\omega_i^3\mu_i^2}{3\pi\varepsilon\hbar c^3},
\end{array}
\eeq

with
\begin{widetext}
\beq\label{secI11}
a(k_0r_{ij}) = \frac{3}{2}\left[ (1-(\hat{\mu}.\hat{r}_{ij})^2)\frac{\sin k_0 r_{ij}}{k_0 r_{ij}}+
(1-3(\hat{\mu}.\hat{r}_{ij})^2) \left[ \frac{\cos k_0 r_{ij}}{(k_0 r_{ij})^2}-\frac{\sin k_0 r_{ij}}{(k_0 r_{ij})^3} \right] \right].
\eeq
\end{widetext}

\noindent The interaction Hamiltonian $H_{int}$ with characteristic frequency $\Omega_{ij}$ governs the coherent part of the evolution and the non-unitary part with characteristic frequency $\gamma_{ij}$ regulates the incoherent part of the evolution. Note that, $\gamma_{ij}$ arises due to multi qubit interaction of the composite system with the bath and is the reason behind the global nature of the
system-bath interaction. Basically, the bath opens up a channel between the two system qubits, an aspect of global interaction not seen for local
interactions. It is to be noted that both $\Omega_{ij}$ and $\gamma_{ij}$ are environment dependent, where the environment factor of $\Omega_{ij}$ is described by the elements in the square brackets of Eq. \eqref{secI8} and the corresponding factor of $\gamma_{ij}$ comes from the parameter $a(k_0r_{ij})$ in Eq. \eqref{secI11} quantifying the coupling between qubits and bath. \\

\noindent
For a two qubit system with identical parties we have $\gamma_{12}=\gamma_{21}$ and $\gamma_1=\gamma_2=\gamma$. Also, $k_0=2\pi/\lambda_0$ is
the resonant wave vector, and occurring in the term $k_0 r_{ij}$ indicates a resonant length scale. Now the damping basis matrix for the equation
(\ref{secI9}) will be

\beq\label{secI12a}
\left(\begin{matrix}
\gamma(N+1)~~0~~\gamma_{12}(N+1)~~0\\
0~~~~~~~~\gamma N~~~~~0~~~~~~~~\gamma_{12}N\\
\gamma_{12}(N+1)~0~~\gamma (N+1)~0\\
0~~~~~\gamma_{12}N~~~~~~0~~~~~~\gamma N\\
\end{matrix}\right).
\eeq

\noindent From (\ref{secI12a}), we see that the condition for positivity is $\gamma_{12} \leq \gamma$, that is, $a(k_0r_{ij})\leq 1$ (for the rest of the paper we will denote it by $a$). Using (\ref{secI11}), we find the specific condition for positivity as

\beq\label{Inew}
R\sin^2 \phi\frac{\sin(k_0r_{ij}-\theta)}{k_0r_{ij}} \leq \frac{2}{3},
\eeq

with

\beq\label{secInew1}
R = \sqrt{\left( 1+\frac{2\cot^2\phi-1}{(k_0r_{ij})^2} \right)^2+\left( \frac{2\cot^2\phi-1}{k_0r_{ij}}\right)^2},
\eeq

and

\beq\label{secInew2}
\tan\theta = \frac{2\cot^2\phi-1}{k_or_{ij}+\frac{2\cot^2\phi-1}{k_0r_{ij}}}.
\eeq

\noindent For a specific case, we can take $\cot\phi=1/\sqrt{2}$, so that $\tan\theta=0$ and $R=1$. For this case, the condition of positivity reduces to

\beq\label{secInew3}
\frac{\sin k_0r_{ij}}{k_0r_{ij}}\leq 1.
\eeq

\noindent From (\ref{secInew3}), it is evident that the limiting condition $(a\rightarrow1)$ can be reached, when the  separation ($r_{ij}$) is very small compared to 
the resonant wavelength $\lambda_0$. This is attainable when $\lambda_0$ is very large, i.e., the separation between the energy levels of both the atomic qubits 
is small. Our aim here is to find a solution for the master equation given by (\ref{secI9}). For that purpose we take a special class of density
matrices of the form of X-states

\beq\label{secI12}
\left(\begin{matrix}
\rho_{11}~~0~~0~~\rho_{14}\\
~0~\rho_{22}~~\rho_{23}~0\\
~0~\rho_{23}^{*}~~\rho_{33}~0\\
\rho_{14}^{*}~~0~~0~~\rho_{44}\\
\end{matrix}\right).
\eeq

\noindent X states are very important in the study of quantum information theory because of their simple representation \cite{x1,x2}. This class includes, among others, the well-known Bell diagonal and Werner states. X states has been studied extensively in the literature \citep{x1,x2}. Their invariance properties and the underlying symmetry have been studied \citep{x2}.  It has been shown that they remain form invariant under the considered quantum operations. Inserting (\ref{secI12}) in (\ref{secI9}), we get the following set of coupled differential equations

\beq\label{secI13}
\begin{array}{ll}
\dot{\rho}_{11} = -2\gamma(N+1)\rho_{11}+\gamma N(\rho_{22}+\rho_{33})+\gamma_{12}N(\rho_{23}+\rho_{23}^{*}), \\
\dot{\rho}_{22} =  \gamma(N+1)\rho_{11}+\gamma N\rho_{44}-\gamma(2N+1)\rho_{22}\\
          ~~~~~~~~-\frac{\gamma_{12}}{2}(2N+1)(\rho_{23}+\rho_{23}^{*})+i\Omega_{12}(\rho_{23}-\rho_{23}^{*}), \\
\dot{\rho}_{33} =  \gamma(N+1)\rho_{11}+\gamma N\rho_{44}-\gamma(2N+1)\rho_{22}\\
          ~~~~~~~~-\frac{\gamma_{12}}{2}(2N+1)(\rho_{23}+\rho_{23}^{*})-i\Omega_{12}(\rho_{23}-\rho_{23}^{*}), \\
\dot{\rho}_{44}=-(\dot{\rho}_{11}+\dot{\rho}_{22}+\dot{\rho}_{33}),\\
\dot{\rho}_{23} = -\gamma(2N+1)\rho_{23}+\gamma_{12}(N+1)\rho_{11}+\gamma_{12}N\rho_{44}\\
          ~~~~~~~~-\frac{\gamma_{12}}{2}(2N+1)(\rho_{22}+\rho_{33})+i\Omega_{12}(\rho_{22}-\rho_{33}),\\
\dot{\rho}_{14} = -\gamma (2N+1)\rho_{14}-4i\omega\rho_{14}.
\end{array}
\eeq

\noindent We are going to solve this for two cases of high and zero temperature. 

\subsection{High temperature case}
For the case when the thermal bath is at high temperature ($N>>1$), the rates of dissipation and absorption are equal. The solution of the master equation is then given by 
\beq\label{secIhotbath}
\begin{array}{ll}
\rho_{11}(t)=W(t)+W'(t),
~\rho_{44}(t)=W(t)-W'(t),\\
\rho_{22}(t)=U(t)+U('t),
~\rho_{33}(t)=U(t)-U'(t),\\
\rho_{23}=V(t)+i V'(t),
\rho_{14}(t) = \rho_{14}(0) \exp \left[-\left( 2\gamma N +4i\omega \right)t\right].
\end{array}
\eeq

where 

\begin{widetext}
\beq\label{secI14a}
\begin{array}{ll}
W(t)=\frac{W(0)+U(0)}{2}-\left(\frac{U(0)-W(0)}{2}\right)e^{-3\gamma Nt}\left[ \cosh \left( \gamma Nt\sqrt{1+8a^2} \right) 
-\frac{\sinh\left( \gamma Nt\sqrt{1+8a^2} \right)}{\sqrt{1+8a^2}}\right] \\
~~~~~~~~+\frac{2a}{\sqrt{1+8a^2}}V(0)e^{-3\gamma Nt}\sinh\left( \gamma Nt\sqrt{1+8a^2} \right),
\end{array}
\eeq

\beq\label{secI14b}
\begin{array}{ll}
U(t)=\frac{W(0)+U(0)}{2}+\left(\frac{U(0)-W(0)}{2}\right)e^{-3\gamma Nt}\left[ \cosh \left( \gamma Nt\sqrt{1+8a^2} \right) 
-\frac{\sinh\left( \gamma Nt\sqrt{1+8a^2} \right)}{\sqrt{1+8a^2}}\right] \\
~~~~~~~~-\frac{2a}{\sqrt{1+8a^2}}V(0)e^{-3\gamma Nt}\sinh\left( \gamma Nt\sqrt{1+8a^2} \right),
\end{array}
\eeq

\beq\label{secI14c}
\begin{array}{ll}
V(t)=V(0)e^{-3\gamma Nt}\left[ \cosh \left( \gamma Nt\sqrt{1+8a^2} \right) +\frac{\sinh\left( \gamma Nt\sqrt{1+8a^2} \right)}{\sqrt{1+8a^2}}\right]
-\frac{2a}{\sqrt{1+8a^2}}(U(0)-W(0))e^{-3\gamma Nt}\sinh\left( \gamma Nt\sqrt{1+8a^2} \right),
\end{array}
\eeq

\beq\label{secInewResub1a}
\begin{array}{ll}
U'(t)=U'(0)e^{-2\gamma Nt}\cos(2\Omega_{12}t)-V'(0)e^{-2\gamma Nt}\sin(2\Omega_{12}t)
\end{array}
\eeq

\beq\label{secInewResub1a}
\begin{array}{ll}
V'(t)=V'(0)e^{-2\gamma Nt}\cos(2\Omega_{12}t)+U'(0)e^{-2\gamma Nt}\sin(2\Omega_{12}t)
\end{array}
\eeq

\beq\label{secInewResub1}
\begin{array}{ll}
 W'(t)=W'(0)e^{-2\gamma Nt}
\end{array}
\eeq

\end{widetext}

\noindent 
From Eq.(\ref{secIhotbath}) and Eq. (\ref{secI14c}), we can clearly see that, due to the presence of global system-environment interaction (whose strength is characterized
by $a$), the off-diagonal components are getting feedback from the diagonal parts and is unlike that of local Markovian decay. If we approximate
that the interactive part of the evolution is negligible, then setting $a \rightarrow 0 $, we get
\beq\label{secI16}
\rho_{23}(t) = \rho_{23}(0)\exp (-2\gamma Nt),
\eeq

which is consistent with the usual local Markovian decay.

\subsection{Zero Temperature case}

Let us now consider the other extreme situation, where the bath is at zero temperature ($N\rightarrow 0$). In this case there will be no absorption part of the Lindbladian. The solution of the master 
equation can then be given by 

\beq\label{secIcoldbath1}
\begin{array}{ll}
\rho_{11}(t)=\rho_{11}(0)e^{-2\gamma t},\\
\rho_{22}(t)=U_c(t)+U_c('t),\\
\rho_{33}(t)=U_c(t)-U'_c(t),\\
\rho_{44}(t)=1-\rho_{11}(t)-2U_c(t),\\
\rho_{23}=V_c(t)+i V'_c(t),\\
\rho_{14}(t) = \rho_{14}(0) \exp \left[-\left( \gamma +4i\omega \right)t\right].
\end{array}
\eeq

where 

\begin{widetext}
\beq\label{secIcoldbath2}
\begin{array}{ll}
U_c(t)=U_c(0)e^{-\gamma t}\cosh (a\gamma t)-V_c(0)e^{-\gamma t}\sinh (a\gamma t)
+\frac{\rho_{11}(0)}{2}\left[ \frac{1+a}{1-a}e^{-(1+a)\gamma t}\left( 1-e^{-(1-a)\gamma t} \right)+\frac{1-a}{1+a}e^{-(1-a)\gamma t}\left( 1-e^{-(1+a)\gamma t} \right) \right],
\end{array}
\eeq

\beq\label{secIcoldbath3}
\begin{array}{ll}
V_c(t)=V_c(0)e^{-\gamma t}\cosh (a\gamma t)-U_c(0)e^{-\gamma t}\sinh (a\gamma t)
+\frac{\rho_{11}(0)}{2}\left[ \frac{1+a}{1-a}e^{-(1+a)\gamma t}\left( 1-e^{-(1-a)\gamma t} \right)-\frac{1-a}{1+a}e^{-(1-a)\gamma t}\left( 1-e^{-(1+a)\gamma t} \right) \right],
\end{array}
\eeq

\beq\label{secIcoldbath4}
\begin{array}{ll}
U'_c(t)=U'_c(0)e^{\gamma t}\cos(2\Omega_{12}t)-V'_c(0)e^{-\gamma t}\sin(2\Omega_{12}t)
\end{array}
\eeq

\beq\label{secIcoldbath5}
\begin{array}{ll}
V'_c(t)=V'_c(0)e^{-\gamma t}\cos(2\Omega_{12}t)+U'_c(0)e^{-\gamma t}\sin(2\Omega_{12}t)
\end{array}
\eeq
\end{widetext}

Here also we can see that for the usual local markovian case, the dynamics of the off-doagonal term is reduced down to $\rho_{23}(t)=\rho_{23}(0)e^{-\gamma t}$.

\section{Evolution of Coherence and entanglement}

Coherence is one of the  fundamental properties of a quantum system closely connected to quantum superposition. Though quantum
optics was the initial framework for understanding the concept and physical significance of coherence \cite{glauber,ecg}, over the years its importance has
been realized in many different fields, such as superconducting devices \cite{almeida} and even in complex biological systems like photosynthetic reaction
centers \cite{lloyd,romero}. Coherence is also very important from the perspective of quantum thermodynamics \cite{new1,new2,new3,new4,new5}. In non-equilibrium situations, presence of
coherence raises serious questions over the classical notion of a thermodynamic bath in a Carnot engine \cite{scully1,scully2}. It has impact on
quantum transport efficiency \cite{mohseni1,huelga,mohseni2}, leading to the violation of Fourier's Law \cite{manzano}. Dissipative quantum
thermodynamics offers the possibility to generate resources which are essential for technologies like quantum communication, cryptography, metrology and
computation \cite{lin}. From these perspectives, it is very important to understand the role of quantum coherence for the future development of robust
quantum memory devices. All these recent developments provided the motivation for constructing a rigorous framework of coherence resource theory \cite{plenio}, where
it was shown that any valid measure of coherence $C(\rho)$ has the following properties:

\begin{enumerate}
  \item $C(\rho)=0$ iff $\rho \in \mathcal{I}$, where $\mathcal{I}$ denotes incoherent states, which are the diagonal states in the preferred basis.
  \item Monotonicity under incoherent selective measurements : $C(\rho) \geq \sum_n p_n C(\rho_n)$. Here $\rho_n=\hat{\mathcal{K}}_n\rho\hat{\mathcal{K}}_n^{\dagger}$ and $p_n=Tr(\hat{\mathcal{K}}_n\rho\hat{\mathcal{K}}_n^{\dagger})$ with $\sum_n \hat{\mathcal{K}}_n^{\dagger}\hat{\mathcal{K}}_n=\mathbb{I}$ and $\hat{\mathcal{K}}_n\mathcal{I}\hat{\mathcal{K}}_n^{\dagger}\subset \mathcal{I}$.
  \item Convexity : $C(\sum_n p_n\rho_n) \leq \sum_n p_n C(\rho_n)$ for any set of states $\{\rho_n\}$ and probability distribution $\{ P_n \}$.
\end{enumerate}

\noindent Based on these properties, the `$l_1$ norm of coherence' and the `relative entropy of coherence' were shown to be valid measures characterizing the coherence
of a quantum system \cite{plenio}. Here, we will take the $l_1$ norm of coherence $C_{l_1}(\rho)$ to study the  dynamics of coherence.
It is an intuitive measure related to the off-diagonal elements of a density matrix and is defined as the $l_1$ matrix norm
$C_{l_1}(\rho)=\sum_{i\neq j}\| \rho_{ij}-\mathcal{I}_{ij}\|$. After doing the optimization over all possible incoherent states ($\mathcal{I}$), it
can be shown to be

\beq\label{secII1}
C_{l_1}(\rho) = \sum_{i\neq j} |\langle i|\rho|j\rangle|,
\eeq

\noindent that is, the sum of the magnitudes of all the off-diagonal elements. Interestingly, it is important to note that the $l_1$-norm of coherence truly captures the notion of 
wave nature as it satisfies a duality relation \cite{arun}. Also it has been shown that the notion of quantum coherence plays a prominent role in understanding of neutrino oscillations
\cite{neutrino}.
For our two cases, we find the $l_1$ norm of coherence to be
\begin{widetext}
\beq\label{secII2}
\begin{array}{ll}
 C_{l_1}^H(\rho) = 2|\rho_{14}(0)|\exp (-2\gamma Nt) +2\sqrt{V^2(t)+V'^2(t)}\\
 C_{l_1}^Z(\rho) = 2|\rho_{14}(0)|\exp (-\gamma t) +2\sqrt{V_c^2(t)+V'^2_c(t)}                    
                     
\end{array}
\eeq
\end{widetext}
For example, let us take a particular Werner state

\beq\label{secII3}
\frac{1}{4}\left(\begin{matrix}
 1-x~~0~~~0~~~0\\
 0~x+1~~-2x~0\\
 0~-2x~~x+1~0\\
 0~~~0~~~0~~1-x\\
\end{matrix}\right),
\eeq

\noindent where $x$ is a non-zero parameter lying between $-1/3$ and 1. For these particular states, the $l_1$ norm of coherence for hot bath is given by Eq. (\ref{secII3a}). Here the coherence defined in the usual computational basis 
 ($\mid00\rangle$, $\mid01\rangle$, $\mid10\rangle$, $\mid11\rangle$), is given by
\begin{widetext}
\beq\label{secII3a}
C_{L1}^H(\rho_W)= xe^{-3N\gamma t} \left[ \cosh \left(\gamma Nt\sqrt{1+8a^2} \right)
+\left(\frac{1+2a}{\sqrt{1+8a^2}}\right)\sinh \left(\gamma Nt\sqrt{1+8a^2} \right) \right].
\eeq
\end{widetext}

The $l_1$ norm of coherence for the zero temperature case is given by 

\begin{widetext}
\beq\label{secII3a1}
\begin{array}{ll}
 C_{L1}^Z(\rho_W)= x e^{-\gamma t}\cosh (a\gamma t)+\frac{1+x}{2}e^{-\gamma t}\sinh (a\gamma t)
-\frac{1-x}{4}\left[ \frac{1+a}{1-a}e^{-(1+a)\gamma t}\left( 1-e^{-(1-a)\gamma t} \right)-\frac{1-a}{1+a}e^{-(1-a)\gamma t}\left( 1-e^{-(1+a)\gamma t} \right) \right]
\end{array}
\eeq
\end{widetext}

\begin{figure}[htb]
	{\centerline{\includegraphics[width=7cm, height=5cm] {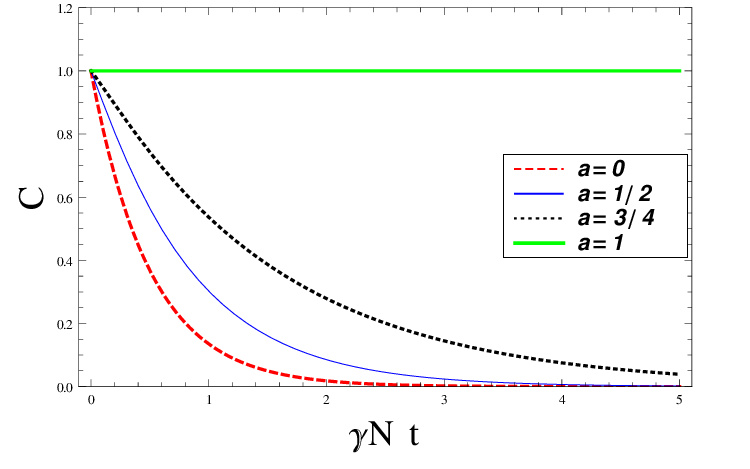}}}
	\caption{(Colour online) Coherence $C$ with respect to $\gamma Nt$ \\
    for the Werner state with $x=1$, with $a$ as a parameter. We have taken $a=0,1/2,3/4,1$ respectively. The red (large dashed) plot is for $a=0$, which is
    the usual Markovian case. The blue (thin line) plot is for $a=1/2$, while black (small dashed) and green (thick line) denote
     $a=3/4$ and $a=1$, respectively. The figure shows that with increment of $a$, the decay of coherence slows down and for the limiting case,
    it freezes after some time. The increment of coherence is the signature of the deviation from Markovian behavior.}
	\label{figVr1}
\end{figure}

In Fig.1, the evolution of $C_{l_1}$ with time (scaled by the decay parameter $\gamma$) is depicted. It can be seen that with the increment of the strength
of the non-local environmental interaction, the decay of coherence slows down. For the
limiting case of $a=1$, the  $l_1$ norm of coherence is constant and equals to $C_{l_1}=x$. This limiting condition is attainable when the separation between the energy levels of 
both the atomic qubits is small. Whereas, for the local Markovian case we have $C_{l_1}=xe^{-2\gamma Nt}$.
Hence it is clear, that the global part of the environmental interaction
imposes a reverse flow of coherence into the system as opposed to the usual Markovian decay. Due to this reverse flow, the usual decoherence process slows down
and may even stop, depending on the strength of the global interaction.
The global interaction therefore generates a feedback to coherence at the expense of the population. In a recent work \citep{frozen1}, a 
dynamical condition has been proposed, under which the coherence of qubit systems is totally unaffected by noise. There the dynamical process under which the 
qubit system evolves is considered to be incoherent; that is, the process maps any incoherent state to the set of incoherent states. From the perspective of 
coherence resource theory, what we are claiming here is different from the motivation of that work. Here, we have taken a dynamical map, which acts as a resource for 
generating quantum coherence in a qubit system. For the low temperature case, we encounter a more interesting situation. Here the dynamics depends on the mixedness $x$ of 
the Werner state we have taken. For the initial pure state charectesized by $x=1$, the Coherence can be given by $C_{L1}^Z=e^{-(1-a)\gamma t}$. Hence, for the limiting case of $a=1$, the coherence is frozen at it's initial maximum value 1. But for other cases ($x<1$), we have a finite contribution from the white noise part of the initial state, which makes the 
situation different. 

\begin{figure}[htb]
	{\centerline{\includegraphics[width=7cm, height=5cm] {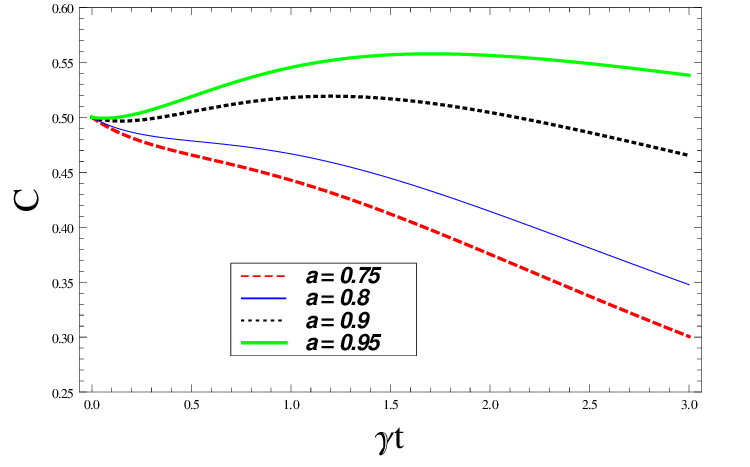}}}
	\caption{(Colour online) Coherence $C$ with respect to $\gamma t$ \\
    for the Werner state with $x=1/2$, with $a$ as a parameter. We have taken $a=0.75,0.8,0.9,0.95$ respectively. The figure shows that with increment of $a$, the decay of coherence 
    slows down and as the global interaction increases, there is a generation of coherence.}
	\label{figVr1}
\end{figure}

From FIG. 2, we see that in the zero temperature regime, with the increase of global interaction, coherence not only stabilizes but also increase from it's initial value, which shows 
a generation of coherence due to the global interaction.

Next, we will consider the dynamics of other quantum correlations like the entanglement and the Fisher information to further 
extend our study. \\

\noindent Let us now compare the dynamics of the coherence with that of the entanglement, expressed as the concurrence \cite{wotters,hill}.
Entanglement is a very basic property of a composite quantum system, which gives rise to nontrivial phenomena in quantum information science. It
is well known that many composite quantum systems having coherence (even some maximally coherent states) may not be entangled at all. For
example, the maximally coherent state $|\psi\rangle = \frac{1}{2}(|00\rangle+|01\rangle+|10\rangle+|11\rangle)$ can be written in the form
$|\psi\rangle=\frac{1}{\sqrt{2}}(|0\rangle+|1\rangle)\otimes\frac{1}{\sqrt{2}}(|0\rangle+|1\rangle)$, that is, the state is a product state and hence
not entangled. This indicates that entanglement is a much more restricted quantum characteristic than coherence. We will now study the
dynamics of concurrence for the global environmental interaction, to see whether it has a similar effect on entanglement as it had on coherence. \\

\noindent For any two qubit system, concurrence may be explicitly calculated from the density matrix as

\beq\label{secII5}
E_c(\rho) = max \{ 0, \sqrt{\lambda_1}-\sqrt{\lambda_2}-\sqrt{\lambda_3}-\sqrt{\lambda_4} \},
\eeq

\noindent where the quantities $\lambda_i$ (i=1,2,3,4) are the eigenvalues of the matrix (in decreasing order )

\beq\label{secII6}
\tau = \rho(\sigma_y\otimes\sigma_y)\rho^{*}(\sigma_y\otimes\sigma_y).
\eeq

\noindent Here $\rho^{*}$ is the complex conjugate of the density matrix $\rho$ in the usual computational basis. For the X states we have taken in (\ref{secI12}),
the concurrence can be expressed in a simpler form as given by

\beq\label{secII7}
E_c = 2max \{ 0, (|\rho_{23}|-\sqrt{\rho_{11}\rho_{44}}), (|\rho_{14}|-\sqrt{\rho_{22}\rho_{33}}) \}.
\eeq

For the case of Werner states (\ref{secII3}), the concurrence will be $E_c = 2max \{ 0, |\rho_{23}|-\sqrt{\rho_{11}\rho_{44}} \}$.
Hence, when $|\rho_{23}| > \sqrt{\rho_{11}\rho_{44}}$, the state is entangled. When the system is interacting with a hot bath, the expression of entanglement for Werner state 
is given by

\begin{widetext}
\beq\label{secII9}
\begin{array}{ll}
E_c^H(\rho_W)=  \frac{x}{2}e^{-3\gamma Nt}\left[3\cosh \left(\gamma Nt\sqrt{1+8a^2} \right)
+\left(\frac{6a-1}{\sqrt{1+8a^2}}\right)\sinh \left(\gamma Nt\sqrt{1+8a^2} \right)\right]-\frac{1}{2}.
\end{array}
\eeq
\end{widetext}

\begin{figure}[htb]
	{\centerline{\includegraphics[width=7cm, height=5cm] {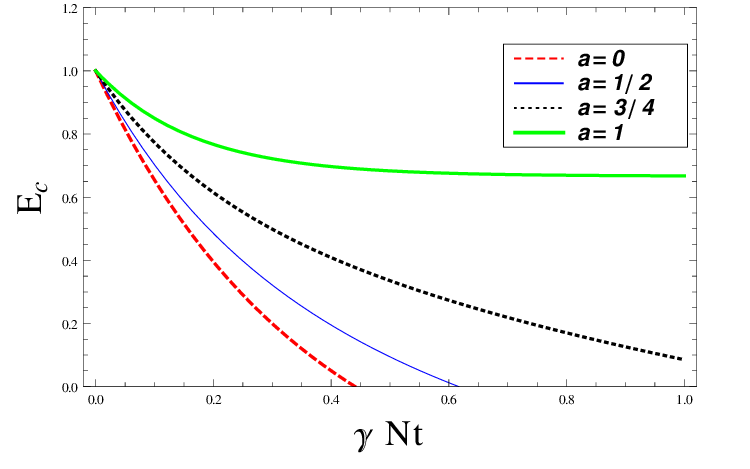}}}
	\caption{(Colour online) Concurrence $E_C$ versus $\gamma t$
    for the Werner state with $x=1$, with $a$ as a parameter. We have taken $a=0,1/2,3/4,1$ respectively. The red (large dashed) plot is for $a=0$, which is
    the usual Markovian case. The blue (thin line) plot is for $a=1/2$, black (small dashed) plot is for $a=3/4$ and green (thick line) is for $a=1$.
    The figure shows that with increment of $a$, the sudden death of entanglement slows down, like a slow decay. For $a=1$,
    the sudden death vanishes completely and entanglement freezes to a particular value after a small initial decay.}
	\label{figVr2}
\end{figure}

\noindent From the Fig. 3, we see that entanglement decay also slows down with the increase in the strength of the global environmental interaction. For the
limiting value $a=1$, after some initial decay, it also saturates like coherence. In this section, we have shown that the global system-bath interaction prolongs the lifetime of 
coherence and entanglement. With the increasing strength of the global part of the interaction (which is characterized by the parameter $a$ ), the lifetime of both the coherence and 
the entanglement increases. \\
Let us now consider the zero temperature situation. If we start from a maximally entangled state ($x=1$), we find that the entanglement can then be given by the expression 
$E(t)=e^{-(1-a)\gamma t}$. Which shows that entanglement decays exponentially. With the increase of the global interaction, the decay slows down and for the limiting
case, it freezes. Similar to the case of coherence, here also the situation is a little different, when we start from an initial mixed state. Let us consider Werner state with $x=1/2$.
In FIG. 4, we can see the evolution of entanglement with time. From the figure we can see that for zero temperature bath, there is a region of the evolution where entanglement is 
generated due to the global nature of the environmental interaction and it even surpasses the initial entanglement.  

\begin{figure}[htb]
	{\centerline{\includegraphics[width=7cm, height=5cm] {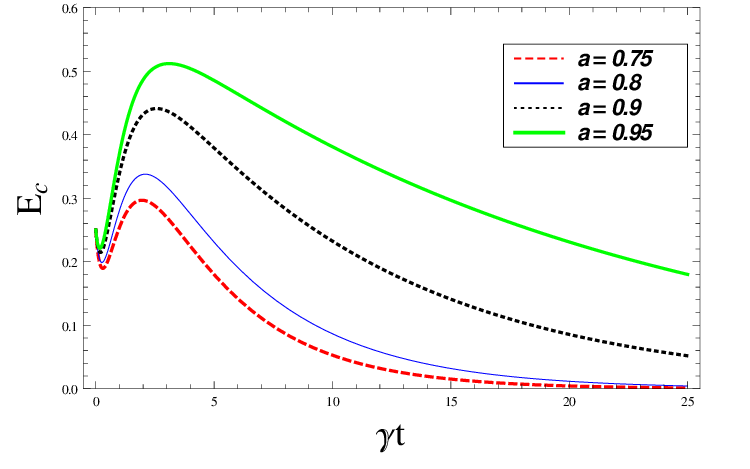}}}
	\caption{(Colour online) Concurrence $E_C$ versus $\gamma t$
    for the Werner state with $x=1/2$, with $a$ as a parameter. We have taken $a=0.75,0.8,0.9,0.95$ respectively.}
	\label{figVr2}
\end{figure}

It is also important to note that not all states will show this behaviour of coherence and correlation freezing. For example, Werner states of the
form $\frac{(1-x)}{4}\mathbb{I}+x|\psi^{+}\rangle\langle\psi^{+}|$ with $|\psi^{+}\rangle=(|00\rangle+|11\rangle)/\sqrt{2}$, will not show the
above observed behavior of slowing down of coherence and correlation decay. This could be attributed to the form of the coherent part of the effective
Hamiltonian (\ref{secI7}) and (\ref{secI7a}), due to the global nature of the system-reservoir interaction. Following Ref. \cite{tanas}, we can infer that under the evolution determined by the interaction Hamiltonian (\ref{secI7a}), the two atom system behaves as a single four level system with the ground state $|g\rangle$, the excited state $|e\rangle$ and two intermediate symmetric and anti-symmetric states $|s\rangle$ and $|as\rangle$, respectively. Where 

\beq\label{N1}
\begin{array}{ll}
|g\rangle = |00\rangle,~|e\rangle = |11\rangle,\\
|s\rangle = \frac{1}{\sqrt{2}}(|10\rangle+|01\rangle),\\
|as\rangle = \frac{1}{\sqrt{2}}(|10\rangle-|01\rangle).
\end{array}
\eeq

The Hamiltonian (\ref{secI7}), which is generated by dipole-dipole interactions \cite{tanas}, does not affect the ground and excited states. Only the symmetric and anti-symmetric 
states are affected by it. The incoherent part of the dipole-dipole interaction is basically the global part of the dynamical map with the strength $\gamma_{12}$. Therefore, the global
interaction affects only  $\rho_{23}$ and $\rho_{23}^*$, which are the components of the symmetric and anti-symmetric states. For the limiting case of $a=1$, the master equation can be written as

\beq\label{newRev}
\begin{array}{ll}
\frac{d\rho}{dt}=\frac{i}{\hbar}[\rho,H]\\
+\gamma(N+1)\left( J^{-}\rho J^{+}-\frac{1}{2}\left(J^{+}J^{-}\rho+\rho J^{+}J^{-}\right)\right)\\
+\gamma N\left( J^{+}\rho J^{-}-\frac{1}{2}\left(J^{-}J^{+}\rho+\rho J^{-}J^{+}\right)\right),
\end{array}
\eeq

with $J^{\pm}=\sigma^{\pm}_1+\sigma^{\pm}_2$. For this case the global interaction creates a decoherence free evolution for $\lvert s\ket$ and $\lvert as\ket$ and thus preserves the non-classicality of the states.

\section{Generation of Quantumness and bound on Fisher information}

In this section, we investigate the generation of non-classicality by the action of global bath in detail. We have considered a measure of quantumness and explicitely find it's connection with quantum Fisher information and quantify the capacity of generation of ``quantumness'' of a global channel in detail. 
The Fisher information has considerable significance in statistical estimation theory, as a measure of ``intrinsic accuracy" \cite{fisher}. A quantum
generalization of the Fisher information was proposed in Refs. \cite{helstrom} and \cite{holevo}. Here, we develop on the theme
that the Fisher information can also be considered as a measure of non-classicality of a quantum system. It has been shown \cite{hall} that the Fisher
information of a quantum observable is proportional to the difference between quantum variance and the classical variance of the conjugate variable. The
Fisher information of a parameterized family of probability densities $\{ p_{\theta} : \theta\in\mathbb{R} \}$ on $\mathbb{R}$, is defined as

\beq\label{secF1}
\begin{array}{ll}
F(p_{\theta}) =\\
 \int_{\mathbb{R}} \left( \frac{\partial}{\partial \theta}p_{\theta}^{1/2}(x) \right)^2 dx
=\frac{1}{4}\int_{\mathbb{R}} \left( \frac{\partial}{\partial \theta}\log p_{\theta}^{1/2}(x) \right)^2p_{\theta} dx.
\end{array}
\eeq

\noindent Particularly, when $P_{\theta}(x)=p_{\theta}(x-\theta)$, then by translational invariance of the Lebesque integral, one can conclude that the Fisher information
$F(p_{\theta})$ is independent of $\theta$. In that case the Fisher information can be denoted as $F(p)$ \cite{luo}. A natural generalization of the Fisher information
\cite{luo, luo1} arises from (\ref{secF1}), when we consider

\beq\label{secF2}
\frac{\partial}{\partial\theta}p_{\theta}=
\frac{1}{2}\left(\frac{\partial}{\partial\theta}\log p_{\theta}.p_{\theta}+p_{\theta}.\frac{\partial}{\partial\theta}\log p_{\theta}  \right).
\eeq

\noindent By replacing the integration by trace, probability $p_{\theta}$ by density matrix $\rho_{\theta}$ and the logarithmic derivative $\frac{\partial}{\partial\theta}\log p_{\theta}$
by the symmetric logarithmic derivative $L_{\theta}$, determined by

\beq\label{secF3}
\frac{\partial}{\partial\theta}\rho_{\theta} = \frac{1}{2}(L_{\theta}\rho_{\theta}+\rho_{\theta}L_{\theta}),~~~~ \theta \in \mathbb{R},
\eeq

\noindent the Fisher information can be expressed as \cite{luo}

\beq\label{secF4}
F(p_{\theta}) = \frac{1}{4}Tr(L_{\theta}^2\rho_{\theta}).
\eeq

\noindent Now, if it is independent of the parameter $\theta$, it can be shown that $F(\rho,H)=\frac{1}{4}Tr(\rho L^2)$ \cite{luo}, where
$i(\rho H-H\rho)=\frac{1}{2}(L\rho+\rho L)$. After some algebra, the Fisher information of an operator $H$ can be shown to be

\beq\label{secF5}
F(\rho,H) = \frac{1}{2}\sum_{m,n}\frac{(\lambda_m-\lambda_n)^2}{\lambda_m+\lambda_n}|\langle\psi_m|H|\psi_n\rangle|^2,
\eeq

\noindent where $\lambda_m$ and $|\psi_m\rangle$ are the eigenvalues and eigenvectors of the density matrix, respectively. It is to be noted that, if $H$ commutes with
$\rho$, then the Fisher information $F(\rho,H)$ becomes zero. Hence, we will have non-zero Fisher information for an observable only when it is skewed or
non-commuting with the density matrix.
In Ref. \cite{hall}, it was shown that, in the Schr\"{o}dinger picture, the position Fisher information is equal to the square of the variance of the
non-classical part of the conjugate momentum with a factor $4/\hbar^2$. This means that the Fisher
information gives us a certain quantification of non-classicality or quantumness of a system. Hence the dynamics of Fisher information would be related to
the dynamics of non-classicality associated with a quantum system. \\
Let us now elaborate the issue of ``non-classicality'' explicitely. In a recent work \cite{vedral}, the notion of quantumness based on the non-commutivity of the algebra of 
observables has been investigated. A measure of quantumness has also been proposed based on the imcompatibility of quantum states. The mutual incompatibility of two given states $\rho_a$
and $\rho_b$ can be quantified by twice of the Hilbert-Schmidt norm of their commutator \cite{srikanth}.

\beq\label{quantumness1}
Q(\rho_a,\rho_b)=2\parallel[\rho_a,\rho_b]\parallel^2 =4 Tr((\rho_a\rho_b)^2-\rho_a^2\rho_b^2)
\eeq

We know that the trace of a positive operator is always positive and vanishes only when the operator is null. So $Q(\rho_a,\rho_b)$ is zero only when the the two density matrix 
commutes with each other. Given this fact, $Q(\rho_a,\rho_b)$ can be considered as a powerfull quantumness witness obeying the relation 

\beq\label{quantumness2}
0 \leq Q(\rho_a,\rho_b) \leq 1
\eeq

The intuitive understanding of $Q(\rho_a,\rho_b)$ as a measure of quantumness is clearly stated in Ref. \cite{vedral}. Given an algebra $\mathbb{A} $ of the system observables, one can 
define the state $\rho$ to be classical if  $Tr (\rho,[A,B])=0~~\forall A,B \in \mathbb{A}$, and the state is quantum otherwise. So the state is classical, if it does not detect the 
non-commutativity of the observables. Hence, intuitively it could be stated that if two states $\rho_a$ and $\rho_b$ do not commute, they are quantum and classical if otherwise. In Ref. \cite{vedral}
it has been shown that $Q(\rho_a,\rho_b)$ is proper witness of the global quantum nature of the given states. Now by chosing $\rho_a=\rho_0$ and $\rho_b=\rho_t$ as the initial and final 
state, we can quantify the generation of quantumness by a certain operation. If $\rho_0$ is a diagonal state in a given basis and the operation only alters the probability distribution,
then the final state will also be a diagonal state. So then there will be no generation of non-commutativity and hence no generation of quantumness by the operation. For example, if we choose the initial 
state to be a incoherent state; i.e. the state is diagonal in the preffered basis and apply only incoherent operation, then the final state will also remain incoherent and no quantumness
will be produced. \\
So let us take the initial state to be a diagonal state $\rho_0=\sum_{j=1}^d \lambda_i \lvert i\ket\bra i\lvert$ and find the quantumness generated by any arbitrary process. Using 
Eq. (\ref{quantumness1}), we find that 

\beq\label{quantumness3}
Q(\rho_0,\rho_t)=2\sum_{i\neq j} (\lambda_i-\lambda_j)^2\lvert\bra i\lvert \rho_t\lvert j\ket\lvert^2
\eeq

Let us now quantify the quantum Fisher information generated by the process starting from a diagonal state. Using Eq. (\ref{secF5}), we find that the generated Fisher information 
can be expressed as 

\beq\label{quantumness4}
F(\rho_0,\rho_t)=\frac{1}{2}\sum_{i\neq j}\frac{(\lambda_i-\lambda_j)^2}{\lambda_i+\lambda_j}|\langle i|\rho_t\lvert j\rangle|^2.
\eeq
Now we know that  

\beq\label{quantumness5}
\frac{1}{\lambda_i+\lambda_j} \geq 1~~~\forall ~~i,j
\eeq

and 

\beq\label{quantumness5a}
|\langle i|\rho_t\lvert j\rangle| \leq 1~~~\forall ~~i,j
\eeq

Using Eq. (\ref{quantumness5}) and (\ref{quantumness5a}), we find the following inequality 

\beq\label{quantumness6}
\frac{Q(\rho_0,\rho_t)}{4} \leq F(\rho_0,\rho_t) \leq \frac{C_{l_1}(\rho_0,\rho_t)}{2}.
\eeq

The relation gives us the intuitive insight on the nature of quantum Fisher information as a measure of quantumness. The  left equality of (\ref{quantumness6}) will strictly hold for two level
system, which is the smallest possible quantum system. For higher dimensional systems and systems consisting more than one party, the situation gets much more complicated. Because for composite
system, quantum correlation comes into picture. From (\ref{quantumness6}), we also surmise that the created coherence is always more than the created Fisher information. \\
Let us now calculate the quantumness and Fisher information generated by our specific global operation. Here we will consider an initial diagonal state of the form 

\beq\label{quantumness7}
\rho_0=\left(\begin{matrix} \rho_{11}(0) & 0 & 0 & 0\\0 & \rho_{22}(0) & 0 & 0\\0 & 0 & \rho_{33}(0) & 0\\0 & 0 & 0 & \rho_{44}(0)\end{matrix}\right)
\eeq

The generated quantumness and Fisher information by the action of our global operation can thus be respectively expressed as 

\beq\label{quantumness8}
Q(\rho_0,\rho_t) = (\rho_{22}(0)-\rho_{33}(0))^2 C^2(\rho_0,\rho_t)
\eeq

\beq\label{quantumness8}
F(\rho_0,\rho_t)=\frac{(\rho_{22}(0)-\rho_{33}(0))^2}{\rho_{22}(0)+\rho_{33}(0)}\frac{C^2(\rho_0,\rho_t)}{4}
\eeq

\noindent where $C(\rho_0,\rho_t)=2\lvert \rho_{23}(t)\lvert$ is the generated coherence. For the case of global environmental interaction, we find the inequality 

\beq\label{quantumness8a}
\frac{Q(\rho_0,\rho_t)}{4}\leq F(\rho_0,\rho_t)\leq \frac{C^2(\rho_0,\rho_t)}{4}.
\eeq

Since $0 \leq C(\rho_0,\rho_t)\leq 1$, we find that the created coherence will always be greater than the created Fisher information and quantumness. 

\noindent For the hot bath, we find the quantumness produced from a diagonal state can be expressed as 

\begin{widetext}
\beq\label{quantumness9}
Q(\rho_0,\rho_t)= 16 U'^2(0)e^{-6\gamma Nt} \left[ \frac{4a^2}{1+8a^2} (U(0)-W(0))^2 \sinh^2\left(\gamma Nt\sqrt{1+8a^2}\right)
 +U'^2(0)e^{2\gamma Nt}\sin^2 (2\Omega_{12}t)  \right]
\eeq
\end{widetext}

The generated quantum Fisher information can be expressed as 

\beq\label{quantumness10}
F(\rho_0,\rho_t)= \frac{Q(\rho_0,\rho_t)}{8U(0)}
\eeq

So we see that the generated Fisher information is equal to the generated quantumness with a factor $1/8U(0)$. Let us consider a particular case with $\rho_{11}(0)=\rho_{44}(0)=0$,
$\rho_{22}(0)=3/4$ and $\rho_{33}(0)=1/4$. For this case $F(\rho_0,\rho_t)=Q(\rho_0,\rho_t)/4$.
Now in the case of high temperature bath ($N>>1$), the contribution from the unitary evolution will not play a significant role. 

\begin{figure}[htb]
	{\centerline{\includegraphics[width=7cm, height=5cm] {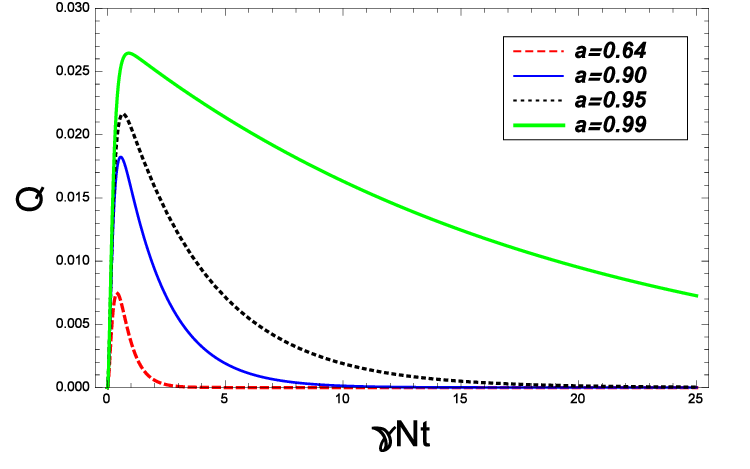}}}
	\caption{(Colour online) Generated Quantumness $Q$ as a function of $\gamma N t$. 
    We have taken the global interaction parameter $a=0.64,0.90,0.95,0.99$ respectively. Corresponding to each value of $a$, we have
    $b=0,-0.45,-0.78,-1.84$  respectively. But for the High temperature case as presented in this plot, the effect of $b$, which is due to the unitary evolution
    is very small}
	\label{figVr3}
\end{figure}

From FIG. 5, we see the generation of quantumness with increasing global interaction. In FIG. 6 we see the generation of coherence with increasing global interaction. The strength of the unitary interaction is characterized by $\Omega_{12}$ given by equation (\ref{secI8}). For the rest of the numerical study, we parametrize $b=\Omega_{12}/\gamma$.

\begin{figure}[htb]
	{\centerline{\includegraphics[width=7cm, height=5cm] {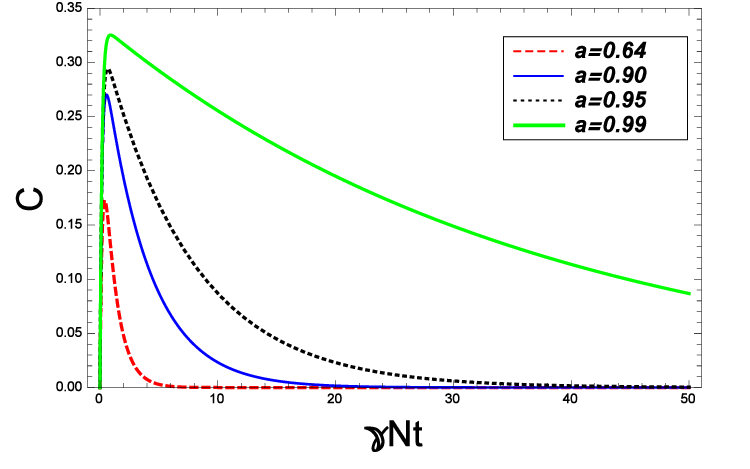}}}
	\caption{(Colour online) Generated Coherence $C$ as a function of $\gamma N t$. 
    We have taken the global interaction parameter $a=0.64,0.90,0.95,0.99$ respectively. Corresponding to each value of $a$, we have
    $b=0,-0.45,-0.78,-1.84$ respectively.}
	\label{figVr3}
\end{figure}

But for the high temperature case, no generation of entanglement is observed. \\
It is important to state that in this case of high temperature bath, the role of the unitary 
evolution in generating quantumness is not prominant, because it is suppressed by the much stronger non-unitary dynamics of the qubit-bath interaction. Whereas in the Zero temperature bath we see a different picture. 

\begin{figure}[htb]
	{\centerline{\includegraphics[width=7cm, height=5cm] {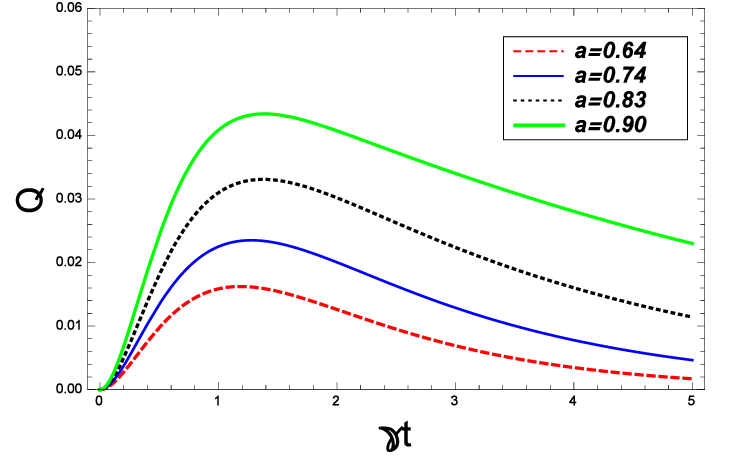}}}
	\caption{(Colour online) Generated Quantumness $Q$ for zero temperature bath case as a function of $\gamma t$. 
    We have taken the global interaction parameter $a=0.64,0.74,0.83,0.90$ respectively. Corresponding to each value of $a$, we have
    $b=0,-0.10,-0.24,-0.45$ respectively.}
	\label{figVr3}
\end{figure}

\begin{figure}[htb]
	{\centerline{\includegraphics[width=7cm, height=5cm] {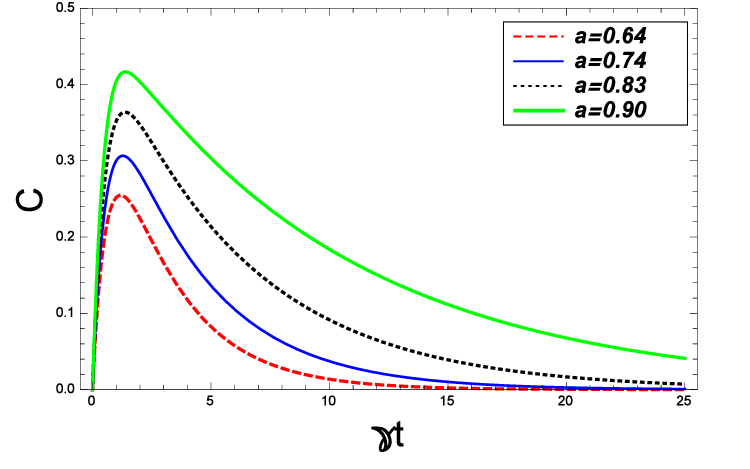}}}
	\caption{(Colour online) Generated Coherence $C$ for zero temperature bath case as a function of $\gamma t$. 
    We have taken the global interaction parameter $a=0.64,0.74,0.83,0.90$ respectively. Corresponding to each value of $a$, we have
    $b=0,-0.10,-0.24,-0.45$ respectively.}
	\label{figVr3}
\end{figure}

\begin{figure}[htb]
	{\centerline{\includegraphics[width=7cm, height=5cm] {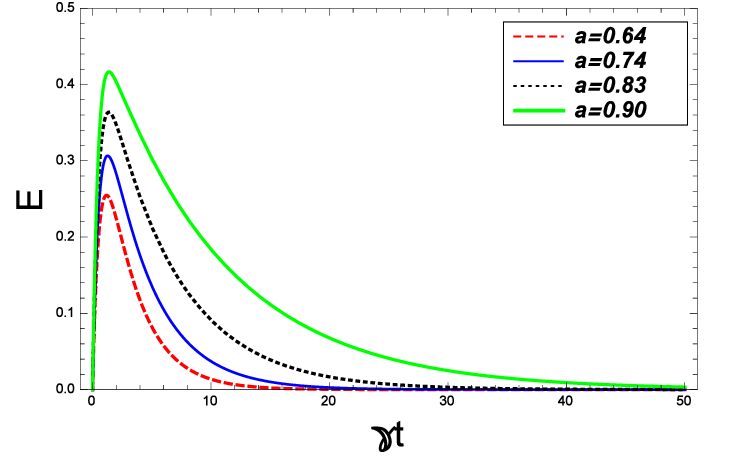}}}
	\caption{(Colour online) Generated Entanglement $E$ for zero temperature bath case as a function of $\gamma t$. 
    We have taken the global interaction parameter $a=0.64,0.74,0.83,0.90$ respectively. Corresponding to each value of $a$, we have
    $b=0,-0.10,-0.24,-0.45$ respectively.}
	\label{figVr3}
\end{figure}

From FIG.7 and FIG.8, we see the generation of quantumness and coherence in zero temperature bath, starting from a initial diagonal state. Importantly, for zero temperature global bath, we can also see a generation of entanglement from FIG. 9, which is unlike the case of high temperature bath.

\section{Global master equation with time varying parameters}

In this section, we further generalize our global master equation by considering the memory effect of the bath. In a practical situation, an
environment usually has memory. In an experiment, a composite quantum system can be exposed to various kind of noises such as vacuum noise, phase noise,
thermal noise as well as a mixture of different kind of noises. Different noise models has been proposed in recent years to model solvable approximate
master equations \cite{Zyczkowski,Daffer}. Correlation dynamics in the Markovian (no memory) regime has been extensively studied \cite{sbindranil}.
However, in practice, an environment is more likely to be non-Markovian. A systematic investigation of the dynamics of quantum coherence and correlation in the presence of
non-Markovian noise is an ongoing study \cite{yu,kubo}. Here we are going to analyze the time dependent global environmental operation based on a non-markovian model of quantum state 
diffusion \cite{gisin}. For certain cases, the Lindblad type master equation can be constructed from non-linear stochastic Schr\"{o}dinger equation of the form 

\beq\label{globalnm1}
\frac{d}{dt} \psi_t=-i H\psi_t +L\psi_t\circ z_t-\frac{1}{2}L^{\dagger}\int_0^{t}\alpha(t,s)\frac{\delta\psi_t}{\delta z_t}ds.
\eeq

where $z_t$ is a white complex valued Wiener process: $z_t=\sum_{\nu}z_{\nu}e^{i\nu t}$. The correlation relations are defined as 

\beq\label{globalnm2}
M[z_t^*z_s]=\delta(t-s)~~;~~M[z_t z_s]=0,
\eeq

and $\alpha(t,s)$ is the environmental correlation function. Here $M[\cdots]$ is the ensemble mean over the classical noise $z_t$ and the system density matrix 
$\rho_t=M[\lvert\psi(t)\ket\bra\psi(t)\lvert]$. The system Hamiltonian is taken to be $H=\frac{\omega}{2}\sigma_z$. The stochastic environmental influence is expressed by the gaussian Wiener process $Z_t$, which drives the system through the operator $L$. For our case of dissipative process described by amplitude damping operation, $L$ is chosen as $\lambda\sigma_{-}$. Here $\lambda$ is a parameter describing the strength of interaction. Now for dissipative interactions, one can choose 

\beq\label{globalnm3}
\frac{\delta\psi_t}{\delta z_t}=f(t,s)\sigma_{-}\psi_t,
\eeq

where $f(t,s)$ is a function to be determined. For the dynamics to be physically consistant this function satisfies the relation \cite{gisin} 

\beq\label{globalnm4}
\partial_t f(t,s)= [i\omega + \lambda F(t)]f(s,t),
\eeq

with the initial condition $f(s,s)=\lambda$. Here $F(t)=\int_0^t\alpha(t,s)f(t,s)ds$. Assuming exponentially decaying environmental correlation 
$\alpha(t,s)=\frac{\gamma_0}{2}\exp\left( -\gamma_0\lvert t-s\lvert +i\xi(t-s) \right)$, we find the reduced map for the system to be 

\beq\label{globalnm5}
\left(\begin{matrix} \rho_{11}(0)e^{-\int_0^t(F(s)+F^*(s))ds} & \rho_{12}(0)e^{-i\omega t-\int_0^tF(s)ds}\\ \rho_{21}(0)e^{i\omega t-\int_0^tF^*(s)ds} & 1-\rho_{11}(t) \end{matrix}\right)
\eeq

with 

\beq\label{globalnm6}
\begin{array}{ll}
F(t)=\frac{\gamma_0}{2\lambda}-\frac{\sqrt{\gamma_0^2-2\gamma_0\lambda^2}}{2\lambda}\\
~~~~~~~~ \times \tanh\left[ \frac{t}{2}\sqrt{\gamma_0^2-2\gamma_0\lambda^2}+\tanh^{-1} \left(\frac{\gamma_0}{\sqrt{\gamma_0^2-2\gamma_0\lambda^2}}\right)\right].
\end{array}
\eeq

Here the resonant situation $\omega=\xi$ is considered. Here two different cases can be considered. The first one is short memory or weak coupling with $\gamma>2\lambda^2$. 
The much more interesting situation appears for long memory or strong coulping case where we have $\gamma<2\lambda^2$. Then we have 

\beq\label{globalnm6}
\begin{array}{ll}
F(t)=\frac{\gamma_0}{2\lambda}+\frac{\sqrt{2\gamma_0\lambda^2-\gamma_0^2}}{2\lambda}\\
~~~~~~~~ \times \tan\left[ \frac{t}{2}\sqrt{2\gamma_0\lambda^2-\gamma_0^2}-\tan^{-1} \left(\frac{\gamma_0}{\sqrt{2\gamma_0\lambda^2-\gamma_0^2}}\right)\right].
\end{array}
\eeq

The master equation of the mentioned operation can be derived from the map given in Eq. (\ref{globalnm5}) with the form 

\beq\label{globalnm7}
\dot{\rho}=i\frac{\omega}{2}[\sigma_Z,\rho]+\lambda[F(t)+F^*(t)]\left(\sigma^{-}\rho\sigma^{+}-\frac{1}{2}\{\sigma^{+}\sigma^{-},\rho\}\right)
\eeq

Note that, here only the zero temperature situation is considered for simplicity. So there is only the dissipation part, but no reheating. But it can always be generalized for 
a finite temperature bath. Now for a two qubit system with identical qubits, where the bath acting globally, we can generalize the master equation as 

\beq\label{globalnm8}
\begin{array}{ll}
\frac{d\rho}{dt} = -\frac{i}{\hbar}[\rho,H_S(t)] 
+\sum_{i,j}\gamma_{ij} (t) \left( \sigma_i^{-}\rho\sigma_j^{+} -\frac{1}{2}\{\sigma_i^{+}\sigma_j^{-},\rho\} \right),
\end{array}
\eeq

with 
\beq\label{globalnm9}
H_S(t)=\sum_{i=1,2}\left(\frac{1}{2}\hbar\sigma_Z^i+\hbar\Omega_{ij}(t)\left(\sigma_i^{-}\sigma_j^{+}+\sigma_i^{+}\sigma_j^{-}\right)\right)
\eeq

and 
\beq\label{globalnm10}
\begin{array}{ll}
\Omega_{12}=\frac{3}{4}\gamma(t)b(k_0r_{12})\\
\gamma_{12}=\frac{3}{2}\gamma(t)a(k_0r_{12})\\
\gamma(t)=\lambda[F(t)+F^*(t)]
\end{array}
\eeq

The solution of Eq. (\ref{globalnm8}) takes the form 

\beq\label{globalnm11}
\begin{array}{ll}
\rho_{11}(t)=\rho_{11}(0)e^{-2G(t)},\\
\rho_{22}(t)=U_n(t)+U_n('t),\\
\rho_{33}(t)=U_n(t)-U'_n(t),\\
\rho_{44}(t)=1-\rho_{11}(t)-2U_n(t),\\
\rho_{23}=V_n(t)+i V'_n(t),\\
\rho_{14}(t) = \rho_{14}(0) \exp \left( -G(t) \right).
\end{array}
\eeq

where 

\begin{widetext}
\beq\label{globalnm12}
\begin{array}{ll}
U_n(t)=U_n(0)e^{-G(t)}\cosh (aG(t))-V_n(0)e^{-(G(t))}\sinh (aG(t))\\
+\frac{\rho_{11}(0)}{2}\left[ \frac{1+a}{1-a}e^{-(1+a)G(t)}\left( 1-e^{-(1-a)G(t)} \right)+\frac{1-a}{1+a}e^{-(1-a)G(t)}\left( 1-e^{-(1+a)G(t)} \right) \right],
\end{array}
\eeq

\end{widetext}

\beq\label{globalnm13}
\begin{array}{ll}
V_n(t)=V_n(0)e^{-G(t)}\cosh (aG(t))-U_n(0)e^{-G(t)}\sinh (aG(t))\\
+\frac{\rho_{11}(0)}{2}\left[ \frac{1+a}{1-a}e^{-(1+a)G(t)}\left( 1-e^{-(1-a)G(t)} \right)-\frac{1-a}{1+a}e^{-(1-a)G(t)}\left( 1-e^{-(1+a)G(t)} \right) \right],
\end{array}
\eeq

\begin{widetext}

\beq\label{globalnm14}
\begin{array}{ll}
U'_n(t)=U'_n(0)e^{-G(t)}\cosh(2bG(t))-V'_n(0)e^{-G(t)}\sinh(2bG(t))
\end{array}
\eeq

\end{widetext}

\begin{widetext}

\beq\label{globalnm15}
\begin{array}{ll}
V'_n(t)=V'_n(0)e^{-G(t)}\cosh(2bG(t))-U'_n(0)e^{-G(t)}\sinh(2bG(t))
\end{array}
\eeq

\end{widetext}

where 
\begin{widetext}
\beq\label{globalnm16}
G(t)=\gamma_0 t 
-2\ln\left[\sqrt{\frac{2\gamma_0\lambda^2}{2\gamma_0\lambda^2-\gamma_0^2}}  \left\lvert\cos\left( \frac{t}{2}\sqrt{2\gamma_0\lambda^2-\gamma_0^2}-\tan^{-1}\left(\frac{\gamma_0}{\sqrt{2\gamma_0\lambda^2-\gamma_0^2}}\right) \right)\right\lvert \right]
\eeq
\end{widetext}

\subsection{Dynamics of coherence and entanglement for Werner state}

In this subsection, we are analysing the dynamics of coherence and concurrence for two qubit Werner state as mentioned in Eq. (\ref{secII3}). For a initial pure state ($x=1$), 
the coherence dynamics is given by $C(\rho_W)=e^{-(1-a)G(t)}$. From FIG. 10 we see that global environmental interaction helps the backflow of information process and as a consequence
we see a periodic revival of coherence with increasing global interaction. 

\begin{figure}[htb]
	{\centerline{\includegraphics[width=7cm, height=5cm] {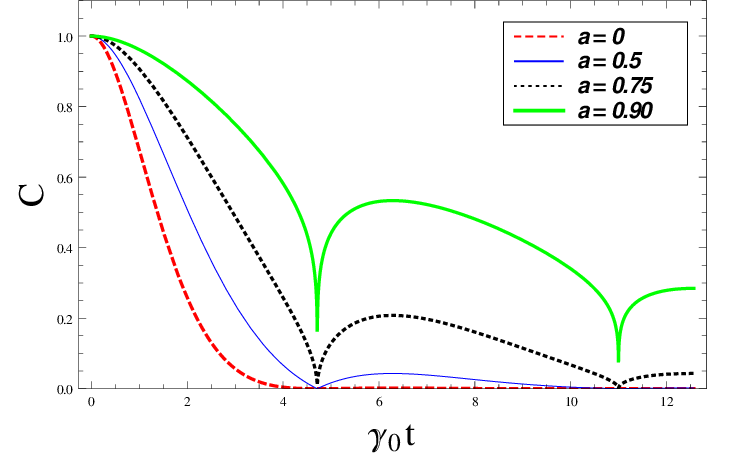}}}
	\caption{(Colour online) Dynamics of coherence $C$ for Werner state with $x=1$ interacting with a zero temperature global non-markovian bath case as a function of $\gamma_0 t$. 
    We have taken the global interaction parameter $a=0,0.5,0.75,0.90$ respectively. Here we have taken $\lambda^2=\gamma_0$.}
	\label{figVr3}
\end{figure}

\begin{figure}[htb]
	{\centerline{\includegraphics[width=7cm, height=5cm] {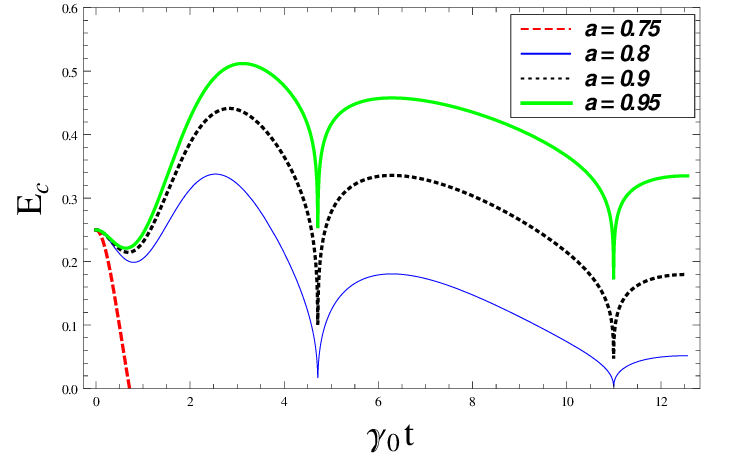}}}
	\caption{(Colour online) Dynamics of entanglement $E_c$ for Werner state with $x=1/2$ interacting with a zero temperature global non-markovian bath case as a function of 
	$\gamma_0 t$. 
    We have taken the global interaction parameter $a=0,0.5,0.75,0.90$ respectively. Here we have taken $\lambda^2=\gamma_0$.}
	\label{figVr3}
\end{figure}

For entanglement also, we see that the global environmental interaction helps the periodic revival of entanglement, as shown in FIG. 11. 

\subsection{Generation of Quantumness}

Here we analyse the generation of quantumness by global non-marokivan interaction. 
Here also we see that global interaction enhances the generation of non-classicality. 

\begin{figure}[htb]
	{\centerline{\includegraphics[width=7cm, height=5cm] {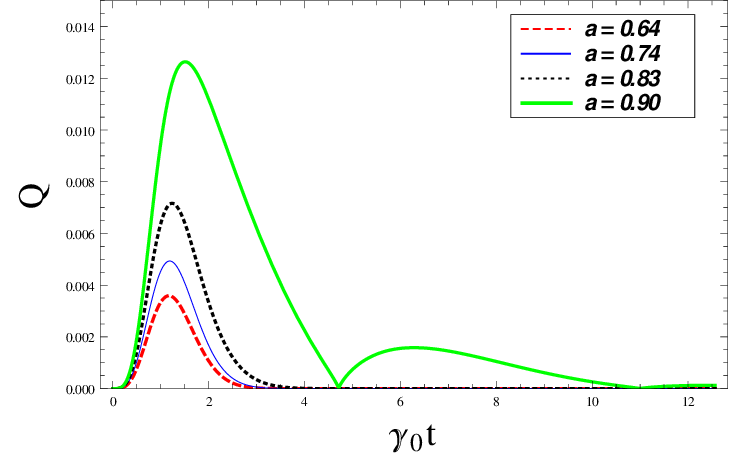}}}
	\caption{(Colour online) Generated Quantumness $Q$ for zero temperature bath case as a function of $\gamma t$. 
    We have taken the global interaction parameter $a=0.64,0.74,0.83,0.90$ respectively. Corresponding to each value of $a$, we have
    $b=\Omega_{12}/2\gamma=0,-0.10,-0.24,-0.45$ respectively.}
	\label{figVr3}
\end{figure}

\section{Conclusion}

To conclude, in this work we exploit a useful global system-environment interaction and study the effect of non-Markovian behaviour on 
various facets of quantum coherence and correlations. Here we have found that the global part of the environmental interaction is acting as a resource to compensate the
decoherence effect. We have further extended our result to the case where the bath has memory.
We have given the exact solution of the proposed master equation for two separate cases of high temperature and zero temperature bath for the memoryless case. For the case for environmental interaction in presence of memory, we have given exact solution for the zero temperature bath. But it can be easily extended to the case of finite temperature bath as well. We have shown that with increasing strength of the global part of the environmental interaction, both coherence and entanglement decay slows down and for the limiting case they eventually freeze. The limiting condition is attainable when the separation between the energy levels of both the atomic qubits is small. For the memory dependent non-Markovian case, we have seen that the global interaction enhances the regeneration of coherence, entanglement and quantumness. Which tells us that the global interaction helps the backflow of information from environment to the system via non-Markovian interaction. It is also very important to mention here that for the case of zero temperature bath, as the strength of the global interaction increases, the unitary interaction between the two qubits will dominate. In that case, the limiting condition is unattainable. Because then the environment cannot resolve the two interacting qubits. Instead it sees the system as a single four level level system in the eigenbasis of it's total Hamiltonian. Moreover, we have also quantified the amount of non-classicality generated by global environmental interaction. It has been shown that the generated quantum Fisher information is lower bounded by the Quantumness with a factor of $1/4$ and upper bounded by the $l_1$ norm of coherence with a factor of 1/2. This gives us an intuitive understanding of quantum Fisher information as a measure of non-classicality. Fisher information and the quantumness measure are both based on the non-commutativity of states and this property is precisely understood as the non-classicality of quantum states. Now from the inequality relation given by Eq. (\ref{quantumness6}), we find that the created coherence by any arbitrary global operation is always greater than the created quantumness and Fisher information. So we understand that coherence contains more than the non-classicality of quantum states and hence cannot be considered as a proper measure of Quantumness. To summarize, in this work we have examined the emergence of non-Markovianity and its impact on the evolution of a number of facets of quantumness in the system. It is observed that non-Markovianity can play a nontrivial and useful role in various quantum information tasks where coherence and entanglement are considered as resources.

\section*{Acknowledgement}

S. Bhattacharya thanks Uttam Singh, Avijit Misra and Titas Chanda of Harish-Chandra Research Institute (HRI) for useful discussions. S. Banerjee
acknowledges the warm hospitality extended to him by the Quantum Information Group at HRI, where this work was initiated. S. Bhattacharya
acknowledges the Department of Atomic Energy, Govt. of India for financial support.

\end{document}